\documentclass[preprint,showpacs,prb,12pt]{revtex4-1}
\usepackage[english]{babel}
\usepackage{amsmath,amssymb} 
\usepackage{times}
\usepackage{epsfig} 
\usepackage{placeins} 

\newcommand{\eq}[1]{\begin{equation}#1\end{equation}}

\newcommand{\tr}{\textrm{Tr}}

\usepackage[colorlinks=true, pdfstartview=FitV, linkcolor=blue,
citecolor=blue, urlcolor=blue]{hyperref} 
\usepackage{hyperref}

\usepackage{color}
\definecolor{nred}   {RGB}{224,0,0}

\begin{document} 

\title{Enhancement of the
thermoelectric power by electronic correlations
in bad metals: a study of the Kelvin formula
}

\author{J. Kokalj$^1$}
\email{jure.kokalj@ijs.si}
\author{Ross H. McKenzie$^2$}
\email{r.mckenzie@uq.edu.au}
\homepage{condensedconcepts.blogspot.com}
\affiliation{$^1$J.\ Stefan Institute, SI-1000 Ljubljana, Slovenia}
\affiliation{$^2$School of Mathematics and Physics, University of Queensland,
  Brisbane, 4072 Queensland, Australia} 
\date{\today}

\begin{abstract}
  In many strongly correlated electron metals the thermoelectric power
  has a non-monotonic temperature dependence and values that are
  orders of magnitude larger than for elemental metals.  Kelvin
  proposed a particularly simple expression for the thermopower in
  terms of the temperature dependence of the chemical potential.  We
  consider a Hubbard model on an anisotropic triangular lattice at
  half filling, a minimal effective Hamiltonian for several classes of
  organic charge transfer salts. The finite temperature Lanczos method
  is used to calculate the temperature dependence of the thermopower
  using the Kelvin formula.  We find that electronic correlations
  significantly enhance the magnitude of the thermopower and lead to a
  non-monotonic temperature dependence.  The latter reflects a
  crossover with increasing temperature from a Fermi liquid to a bad
  metal. Although, the Kelvin formula gives a semi-quantitative
  description of some experimental results it cannot describe the
  directional dependence of the sign of the thermopower in some
  materials.
\end{abstract}

\pacs{71.72.+a, 71.30.+h, 74.25.-q, 74.70.Kn, 75.20.-g}
\maketitle 

\section{Introduction.}

\begin{figure}[htb] 
 \centering 
\includegraphics[width =92mm,angle=-90] {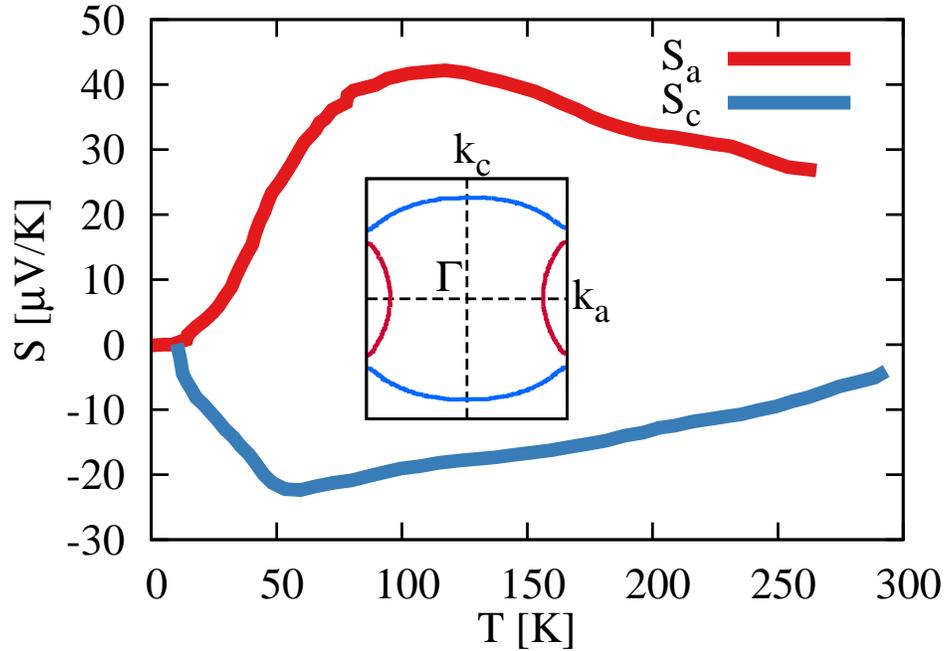}
  \caption{(color online) Temperature
dependence of the thermoelectric power in
the organic metal $\kappa$-(BEDT-TTF)$_2$Cu[N(CN)$_2$]Br.
The two different curves correspond to two different directions in the crystal.
Experimental data is taken from Ref. \onlinecite{yu91}.
Note  the non-monotonic temperature dependence and that
the thermopower is comparable to $k_B/e =  86 \ \mu$V/K.
For temperatures below about 50 K the thermopower is approximately linear 
in temperature,
as expected in a Fermi liquid.
The inset shows a schematic of the electron and hole Fermi surfaces
deduced from a tight-binding band structure \cite{merino00b}.
Transport in the $a$ and $c$ directions will be dominated by holes
and electrons, respectively.
}
\label{fig_1}
\end{figure}

Strongly correlated electron materials have attracted interest as
candidate thermoelectric materials because they can exhibit values of
the Seebeck coefficient $S$ as large as 100 $\mu$V/K.\cite{zlatic09}
 Understanding
and describing the temperature dependence of $S$ in strongly correlated
materials represents a significant theoretical challenge. Both the
magnitude and the temperature dependence of $S$ is distinctly
different than in elemental metals. At low temperatures $S$ increases
linearly with temperature, with a large slope, and reaches a maximum
value of order $k_B/e \simeq $ 86 $\mu$V/K ($k_B$ is Boltzmann’s
constant and $e$ is the charge of an electron). With increasing
temperature $S$ decreases and can even change sign. 
These qualitative features are seen in diverse materials
including organic charge transfer salts \cite{yu91}, cuprates \cite{obertelli92,honma}, heavy fermion compounds \cite{mun12}, cobaltates \cite{wang03},
and iron pnictides \cite{hodovanets13}.
This is illustrated in Figure \ref{fig_1} with experimental results
for an organic metal.
Behnia, Jaccard, and Floquet showed that
for a wide range of materials that the slope of the temperature
dependence of $S$ and the specific heat capacity at low temperatures
were proportional to one another.\cite{behnia04} For heavy fermion
materials, this observation can be explained in terms of
a slave boson treatment of the Kondo lattice model.\cite{zlatic08}

Understanding the thermopower in strongly correlated electron
materials has recently attracted increasing theoretical interest.\cite{zlatic09,tomczak12}
Shastry and coworkers have argued \cite{peterson07,peterson07a,
  shastry09} that the high frequency limit of the 
Kubo formula for the thermopower 
actually gives a good approximate value to the dc limit. This approach
has the
advantage that the thermopower (a transport property) can actually be
evaluated from an equal-time expectation value (an equilibrium
property). Peterson and Shastry \cite{peterson10} have shown that the
thermopower is approximately given by the Kelvin formula, the
derivative of the entropy with respect to the particle number; which
via a thermodynamic Maxwell identity equals the derivative of the
chemical potential with respect to temperature.  
Recent Dynamical Mean-Field Theory (DMFT) calculations for the Hubbard
model\cite{arsenault} and the Falicov-Kimball\cite{zlatic14} model
show that in the bad metal regime the Kelvin formula is a reasonable
approximation.
The Kelvin formula has the significant advantage that a transport
property can be calculated from an equilibrium thermodynamic
property. It also illuminates the physical significance of work by
Jakli\v{c} and Prelov\v{s}ek who showed\cite{jaklic96} that for the
$t-J$ model the entropy as a 
function of doping is a maximum close to optimal doping.  This means
that the thermopower should change sign at optimal doping, as is
observed experimentally in the cuprates \cite{obertelli92,honma}.

Figure \ref{fig_1} shows the measured temperature dependence of
the thermopower of an organic metal.\cite{yu91}
The authors also calculated the thermopower using a Boltzmann equation and
a band structure
obtained with the Huckel approximation. They obtained values that
were about five times smaller than the experiment. However, they found that
if all the hopping integrals were reduced by about a factor of five that
the results were comparable to experiment. 
Similar results were obtained earlier  by Mori and Inokuchi \cite{mori88}.
Merino and McKenzie suggested that the non-monotonic temperature
dependence arose from a crossover with increasing temperature from
a renormalised Fermi liquid to a bad metal \cite{merino00}.
They showed this was consistent with results of calculations
for a Hubbard model
based on Dynamical Mean Field Theory.

\section{The Kelvin formula}

Starting from a Kubo formula Peterson and Shastry showed that if
one interchanges the thermodynamic and the static limits that the thermopower is given
by the temperature derivative of the chemical potential \cite{peterson10} 
\begin{equation}
S_K =
-{1 \over e}\left({\partial \tilde S \over \partial N_{el}} \right)_{T,V}=
{1 \over e}\left({\partial \mu \over \partial T} \right)_{N,V} ,
\label{kelvin}
\end{equation}
where $e$ is the magnitude of the charge of an electron, $\tilde S$ is the entropy, and $N_{el}$ is the particle number. 
Note that this result is independent of the direction of the thermal gradient 
in the crystal.
Hence, it will be unable to explain the origin of the different
signs shown in Figure \ref{fig_1}.
As a result of the third law of thermodynamics, the entropy should vanish
as the temperature goes to zero for all $N_{el}$ and so $S_K(T) \to 0$ as $T \to 0$.

\section{Hubbard model}

For numerical calculations we consider a system at fixed temperature
$T$ and chemical potential $\mu$, and model it with a (grand
canonical) Hubbard model on the anisotropic triangular lattice,
\begin{equation}
  \hat H_{el}=-\sum_{i,j,s} t_{i,j} c_{i,s}^\dagger c_{j,s} +U\sum_i
  \hat n_{i,\uparrow} \hat n_{i,\downarrow} -\mu \hat N_{el}.
\label{eq_hel}
\end{equation}
This is a minimal effective Hamiltonian for several
classes of organic 
charge transfer salts \cite{powell11} when at half filling.
$\hat N_{el} \equiv \sum_{i,s} \hat n_{i,s}$
is the total electron number operator,
$t_{i,j}=t$ for nearest neighbour bonds in two directions and
$t_{i,j}=t'$ for nearest neighbour bonds in the third
direction. Electronic spin is denoted with $s$ ($\uparrow$ or
$\downarrow$).  

For fixed half filled system the chemical potential changes with
temperature and $\mu(T)$ is fixed by the constraint that 
\begin{equation}
 \langle \hat N_{el} \rangle = N ,
\end{equation}
where $N$ is the number of lattice sites, ensuring half-filling, and 
$\langle \hat A \rangle$ denotes the grand canonical thermal average,
$\langle \hat A \rangle \equiv \tr [\hat A \exp(-\beta \hat H_{el})]/Z$ with $Z$
being the thermodynamic sum $Z=\tr [\exp(-\beta \hat H_{el})]$. Here we
have also used $\beta=1/(k_B
T)$.

Our numerical results for finite lattices were obtained by the
finite-temperature Lanczos method (FLTM) \cite{jaklic00}, which 
we previously used to determine several thermodynamic quantities of
this Hubbard model \cite{kokalj13}. We showed that
there was a transition from a metal to a Mott insulator with increasing $U/t$,
with the critical value depending on the amount of frustration $t'/t$.
In the metallic phase as the temperature increased there is a crossover
from a Fermi liquid (with a specific heat and entropy that increased linearly with
temperature) to a bad metal, characterised by an entropy of order $k_B \ln (2)$.
The coherence temperature associated with this crossover, was substantially
reduced by strong correlations, having a value of order $t/10$.

\section{Results}

\begin{figure}[htb] 
 \centering 
\includegraphics[width
=92mm,angle=-90]{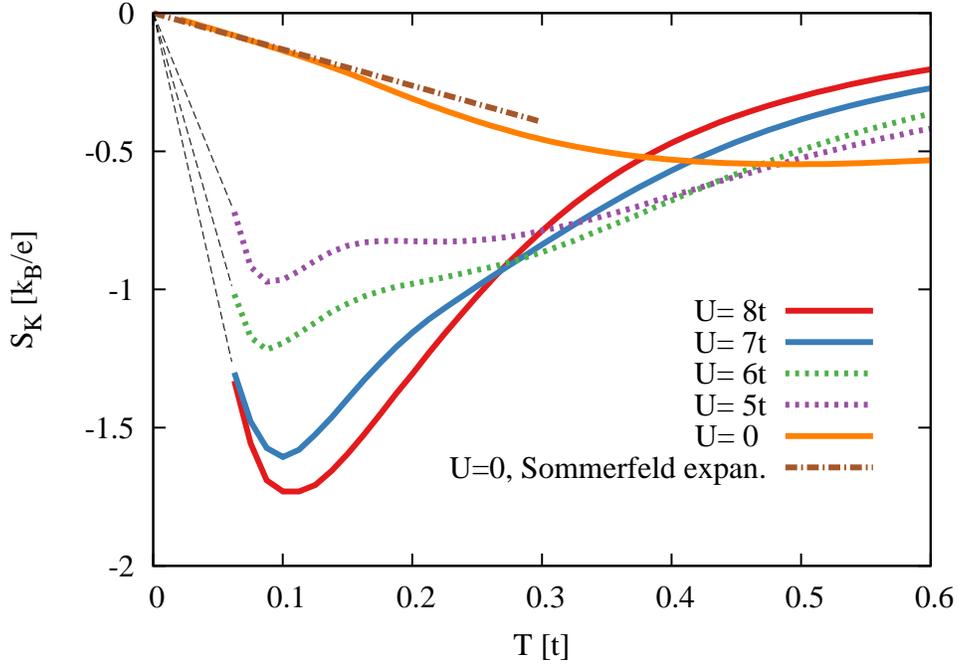}  
\caption{(color online) Enhancement of the thermopower by strong
  correlations.  The temperature dependence of the Kelvin thermopower
  $S_K$ is shown for several different values of the Hubbard $U$.  All
  results are for the isotropic triangular lattice, $t'=t$.  As the
  Mott metal-insulator transition ($U_c \simeq 7.5 t$)\cite{kokalj13}
  is approached the magnitude of the thermopower increases to values
  that are an order of magnitude larger than for non-interacting
  electrons ($U=0$) for temperatures of about $T \sim t/10$. The
  maximum in $|S_K|$ at low temperatures corresponds to the crossover
  from a Fermi liquid at low temperatures to a bad metal at higher
  temperatures.  This maximum is also seen in the specific heat
  \cite{kokalj13} and the spin susceptibility.  The curves have been
  linearly extrapolated from their value at $T=0.06t$ to zero at
  zero-temperature.  Also shown is the linear temperature dependence
  obtained by a Sommerfeld expansion for non-interacting electrons
  \cite{Ashcroft}.  }
\label{fig_2}
\end{figure}

In Fig. \ref{fig_2} we show how the thermopower estimated with Kelvin
formula $S_K$ shows a large enhancement with increasing electronic
interactions $U$ at low $T$. In comparison to the non-interacting ($U=0$)
system the enhancement can be  an order of magnitude and originates
in electronic correlations. The largest magnitude of $S_K$ is reached for
$T\sim T_\textrm{coh}\sim 0.1t$ which is much lower than the Fermi energy. Below
$T_\textrm{coh}$ one enters a coherent Fermi liquid regime in which one
expects a linear temperature dependence of $S_K$, extrapolating
to zero at zero temperature, in accordance with the third law
of thermodynamics. This regime is hard to
reach numerically and our results only indicate it with $S_K$ tending
to $0$ at $T\to 0$ for $T<T_\textrm{coh}$. In Fig. \ref{fig_2} we linearly
extrapolated $S_K$ to 0 for $T\to 0$ by hand to demonstrate the expected
behaviour. 

\subsection{Non-interacting fermions}
\begin{figure}[htb] 
 \centering 
\includegraphics[width =92mm,angle=-90]{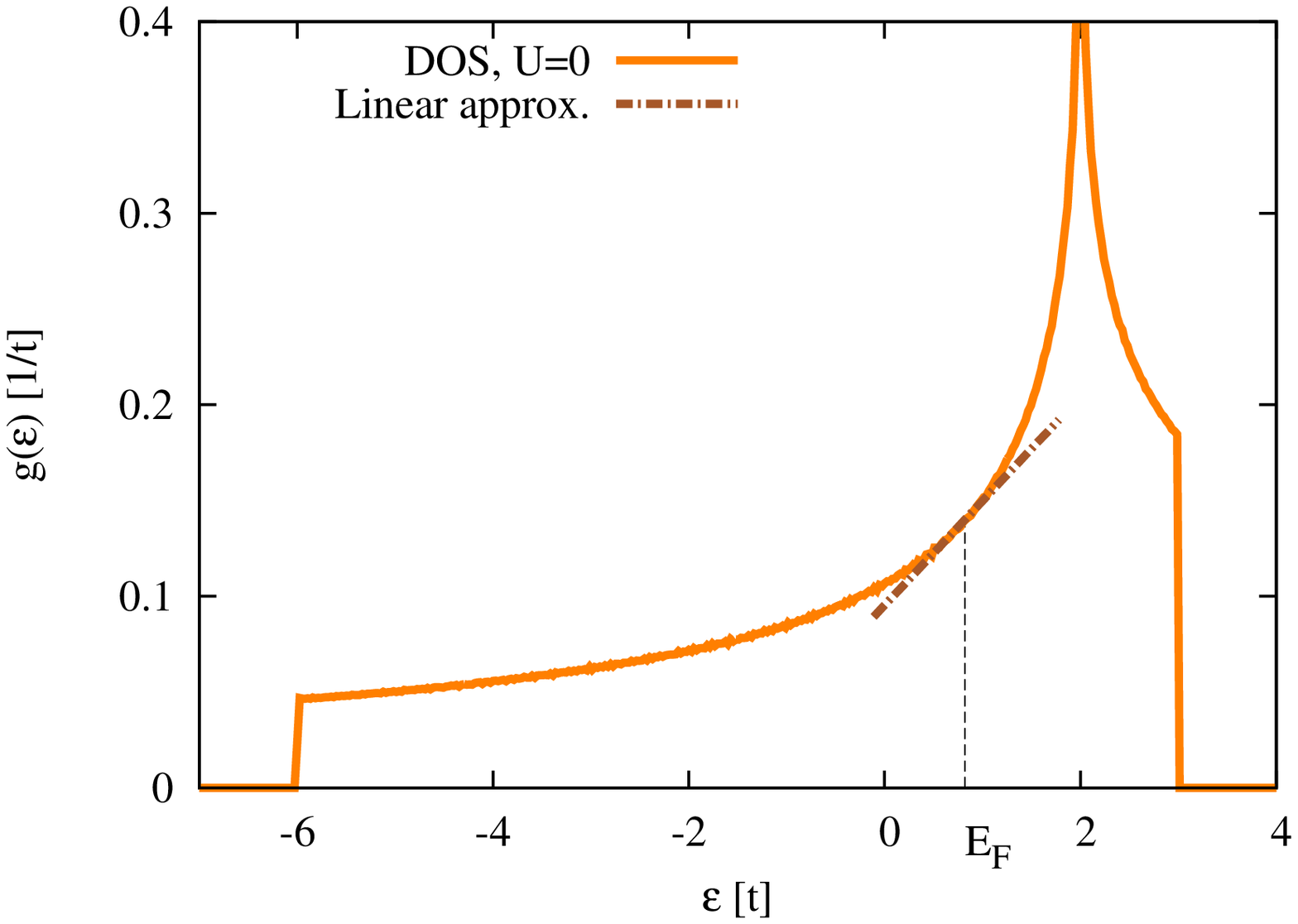}
\caption{(color online) Energy dependence of the density of states
  $g(\epsilon)$ for the tight-binding band structure associated with
  non-interacting electrons with $t'=t$. For half filling $\mu = E_F =
  0.82t$ at zero temperature, and $g(E_F) = 0.14/t$ and $g'(E_F) =
  0.056/t^2$.  The latter determines the slope of the Kelvin
  thermopower versus temperature for non-interacting
  electrons. Reversing the sign of $t'$ or both $t$ and $t'$,
  corresponds to a particle-hole transformation and reverses the sign
  of this derivative ($g'(E_F)$).  }
\label{fig_3}
\end{figure}

In a non-interacting fermion system the chemical potential, at
temperatures much less than the Fermi temperature can be estimated via
Sommerfeld expansion leading to \cite{Ashcroft}
\begin{equation}
\mu(T) = E_F - {\pi^2 \over 6}(k_B T)^2 {g'(E_F) \over g(E_F)}.
\label{ashcroft-mu}
\end{equation} 
Here $g(E_F)$ is the density of states (DOS) at the Fermi energy
($E_F$) and $g'(E_F)$ is its slope. Substituting
Eq. (\ref{ashcroft-mu}) in the Kelvin formula gives $S_K$ that is
linear in temperature, with a magnitude of order, $(k_B/e)(k_B
T/E_F)$, which for elemental metals will be very small. We show in
Fig. \ref{fig_2} that the Sommerfeld expansion,
Eq. \eqref{ashcroft-mu}, gives a good low $T$ estimate for
non-interacting electrons, up to about $T=0.3t$.  The non-interacting
density of states is shown in Fig. \ref{fig_3}.


\subsection{Fermi liquid regime}
\label{sec_fl}

When using the Kelvin formula one should however be careful, since it
may not be a good approximation in some regimes. For example, its
weakness for $T<T_\textrm{coh}$ can be understood by starting with the Mott
formula \cite{shastry09,peterson10}
\eq{
S_{Mott}=-T\frac{\pi^2k_B^2}{3e}\frac{d}{d\mu}
\ln[
g(\mu)\overline{v_{k,x}^2 \tau_{k,\mu}}
]|_{\mu\to E_F}.
\label{eq_smott}
}  
Here, $\overline{v_{k,x}^2 \tau_{k,\mu}}$ denotes the average of the
quasi-particle velocity at wave vector $k$ in the $x$ direction
($v_{k,x}$) times the quasi-particle lifetime ($\tau_{k,\mu}$) over a
surface in reciprocal space at energy equal to $\mu$.  To obtain the
Kelvin formula from $S_{Mott}$ one needs to neglect the $\mu$
dependence of $\overline{v_{k,x}^2 \tau_{k,\mu}}$ in
Eq. \eqref{eq_smott} leading to
\eq{
S_{K}=-T\frac{\pi^2k_B^2}{3e}\frac{d}{d\mu}
\ln[
g(\mu)
]|_{\mu\to E_F}.
\label{eq_sklowt}
}  

This is the same result as obtained for the non-interacting case via
Eq. \eqref{ashcroft-mu} and also represents the low-temperature Kelvin
formula in a coherent regime with well defined quasi-particles.  The
problem with the Kelvin formula in a Fermi liquid regime is in
neglecting the $\mu$ dependence of the velocity in the term $\overline{v_{k,x}^2
  \tau_{k,\mu}}$, while keeping the $\mu$ dependence of the density of states $g$, which is also
related to the velocity since $g\propto 1/v$. It is also unlikely that in a Fermi liquid regime that
$\tau$ would  cancel the $\mu$ dependence of
$v_{k,x}^2$ in $\overline{v_{k,x}^2 \tau_{k,\mu}}$.  That the Kelvin
formula is more appropriate for higher temperatures and in the incoherent
regime was already pointed out in Ref. \onlinecite{peterson10}, while in the
low temperature regime it only gives a rough approximation.
This is explicitly found in
recent Dynamical Mean-Field Theory (DMFT) calculations for the Hubbard
model\cite{arsenault} and the Falicov-Kimball\cite{zlatic14} model.

\subsection{Effect of dimerization}
\label{sec_dimer}

In Fig. \ref{fig_1} the measured thermopower of an organic metal in
two different directions is shown.  The opposite signs for the
two directions was argued\cite{yu91} to originate in the finite
dimerization of the hopping (alternating hopping $t-\delta t$,
$t+\delta t$, \ldots) in two directions on the triangular
lattice. Such a dimerisation splits the band into two bands, one
electron and the other hole like \cite{ivanov97,merino00b}. Each band
dominates the thermopower in its own direction and leads to opposite
signs of the thermopower for the two directions.  Due to the band
splitting the density of states is also split. However, it turns out
that just the density of states cannot capture the change of sign and
that $v^2$ term discussed in Sec. \ref{sec_fl} needs to be included to
reproduce the opposite signs. The Boltzmann transport equation
approach in Ref. \onlinecite{yu91} does take these terms into account
and captures the correct signs.

\subsection{Comparison to experiment}

For comparison of the experimental data shown in Fig. \ref{fig_1} and
our results shown in Fig. \ref{fig_2} we set the energy scale $t=50$
meV $\sim 580$ K as appropriate value obtained by Density Functional
Theory for organic charge transfer
salts \cite{kandpal09, nakamura09, jeschke12,scriven12}. 
We note that with our definition of hopping parameters in
Eq.~\eqref{eq_hel} we should for organics either take both $t$ and
$t'$ negative\cite{kandpal09,jeschke12,scriven12} or positive $t$ and
negative $t'$\cite{nakamura09}, but both changes correspond at
half-filling to particle-hole transformation (with additional shift in
$k$ space for the later) and therefore only reverse the sign of $S_K$
shown in Fig. \ref{fig_2}.
Then we
estimate from Fig. \ref{fig_2} that the maximal thermopower would
appear at roughly $T_\textrm{coh}=60$ K, which is in agreement with
experiment. We also capture the qualitative $T$ dependence of the
thermopower. However, as already discussed in Sec. \ref{sec_dimer},
the Kelvin formula does not have the potential to describe the
orientational dependence shown in Fig. \ref{fig_1}, which originates
in the finite dimerization of the lattice.

\section{Conclusion}

In conclusion, we have shown with the Kelvin forumla, which is a good
approximation in the bad metallic regime, that the thermopower is
strongly enhanced by electronic correlations at low $T$, even by an
order of magnitude compared to the weak or non-interacting electron
limit.  Comparing with experimental data for an organic charge
transfer salt, we capture qualitatively the temperature dependence and
overall magnitude of the thermopower.  On the other hand, the Kelvin
formula can not capture the orientational dependence of $S$ observed
in experiment, for which one would need to employ a Kubo formula and
introduce dimerization of the lattice into the model.  We leave this
as a future challenge.

\begin{acknowledgments}
  We acknowledge helpful discussions with Nandan Pakhira, Peter
  Prelov\v{s}ek, Sriram Shastry and Jernej Mravlje. 
  This work was supported by
  Slovenian Research Agency grant Z1-5442 (J.K.) and an Australian
  Research Council Discovery Project grant.
\end{acknowledgments}

\bibliographystyle{apsrev4-1}
\bibliography{ref_kelvin}

\end{document}